\documentclass{article}
\usepackage{tabularx}
\usepackage{booktabs}
\usepackage{ismir,cite}
\usepackage{graphicx}
\usepackage{color}
\usepackage[utf8]{inputenc}
\usepackage{url}
\usepackage{graphicx}
\graphicspath{{/Users/ecreager/git/fmnmf/images/}}
\usepackage{tikz}
\usepackage{epstopdf}
\usetikzlibrary{arrows}
\usepackage{amsmath ,amssymb}
\usepackage[labelformat=simple]{subcaption}

\title{Nonnegative tensor factorization with frequency modulation cues for blind audio source separation}

\multauthor
{Elliot Creager$^{1,3}$ \hspace{1cm} Noah D. Stein$^1$ \hspace{1cm} Roland Badeau$^{2, 3}$\hspace{1cm} Philippe Depalle$^{3}$} {
  $^1$ Analog Devices Lyric Labs, Cambridge, MA, USA\\
$^2$ LTCI, CNRS, T{é}l{é}com ParisTech, Universit{é} Paris-Saclay, Paris, France\\
$^3$ CIRMMT, McGill University, Montr{é}al, Canada\\
\tt\small elliot.creager@analog.com, noah.stein@analog.com, \\ \tt\small roland.badeau@telecom-paristech.fr, depalle@music.mcgill.ca
}
\def\authorname{Elliot Creager, Noah D. Stein, Roland Badeau, Philippe Depalle}

\begin{document}
\newcommand{\hop}{\mathcal{H}} 
\newcommand{\X}{\mathcal{X}} 
\newcommand{\p}{p^{\text{obs}}(f, t, \fsfr)} 
\newcommand{\q}{q(f, t, \fsfr)} 
\newcommand{\qr}{q(\fsfr | s, t)} 
\newcommand{\qf}{q(f | z, s)} 
\newcommand{\qt}{q(z, t | s)} 
\newcommand{\qrd}{q(\fsfr | s, t-1)} 
\newcommand{\qrfirst}{q(\fsfr | s, t=1)} 

\newcommand{\qq}{q(f, t)} 
\newcommand{\qqj}{q(f, t, z)} 
\newcommand{\qqw}{q(f| z)} 
\newcommand{\qqh}{q(z, t)} 
\newcommand{\pp}{p(f, t)} 
\newcommand{\ppj}{p(f, t, z)} 

\newcommand{\fsfr}{r} 
\newcommand{\FSFR}{R}
\maketitle
\begin{abstract}\label{sec:abstract}
We present Vibrato Nonnegative Tensor Factorization, an algorithm for single-channel unsupervised audio source separation with an application to separating instrumental or vocal sources with nonstationary pitch from music recordings.
Our approach extends Nonnegative Matrix Factorization for audio modeling by including local estimates of frequency modulation as cues in the separation.
This permits the modeling and unsupervised separation of vibrato or glissando musical sources, which is not possible with the basic matrix factorization formulation.

The algorithm factorizes a sparse nonnegative tensor comprising the audio spectrogram and local frequency-slope-to-frequency ratios, which are estimated at each time-frequency bin using the Distributed Derivative Method.
The use of local frequency modulations as separation cues is motivated by the principle of common fate partial grouping from Auditory Scene Analysis, which hypothesizes that each latent source in a mixture is characterized perceptually by coherent frequency and amplitude modulations shared by its component partials.
We derive multiplicative factor updates by Minorization-Maximization, which guarantees convergence to a local optimum by iteration.
We then compare our method to the baseline on two separation tasks: one considers synthetic vibrato notes, while the other considers vibrato string instrument recordings.
\end{abstract}
\section{Introduction}\label{sec:intro}
Nonnegative matrix factorization (NMF) \cite{lee99} is a popular method for the analysis of audio spectrograms \cite{smaragdis03}, especially for audio source separation \cite{smaragdis14}.
NMF models the observed spectrogram as a weighted sum of rank-1 latent components, each of which factorizes as the outer product of a pair of vectors representing the constituent frequencies and onset regions for some significant component in the mixture, e.g. a musical note.
Equivalently, the entire spectrogram matrix approximately factorizes as a matrix of spectral templates times a matrix of temporal activations, typically such that the approximate factors have many fewer elements than the full observation.
While NMF can be used for supervised source separation tasks with a straightforward extension of the signal model \cite{smaragdis07}, this necessitates pre-training NMF representations for each source of interest.

The use of modulation cues in source separation is popular in the Computational Auditory Scene Analysis (CASA) \cite{wang06} literature, which, unlike NMF, typically relies on partial tracking.
E.g., \cite{wang95} isolates individual partials by frequency warping and filtering, while \cite{li09} groups partials via correlations in amplitude modulations.
\cite{barker13}, which more closely resembles our work in the sense of being data-driven, factorizes a tensor encoding amplitude modulations for speech separation.

Our approach is inspired by \cite{stein15} and \cite{traa15}, which present a Nonnegative Tensor Factorization (NTF) incorporating direction-of-arrival (DOA) estimates in an unsupervised speech source separation task.
Whereas use of DOA information in that work necessitates multi-microphone data, we address the single-channel case by incorporating the local frequency modulation (FM) cues at each time-frequency bin.
These cues are combined with the spectrogram as a sparse observation tensor, which we factorize in a probabilistic framework.
The modulation cues are adopted structurally by way of an NTF where each source in the mixture is modeled via an NMF factor and a time-varying FM factor.
\section{Background}\label{sec:background}
\tikzset{observed node/.style={circle,minimum size=0.8cm,fill=gray!30,draw,font=\sffamily\large\bfseries},
unobserved node/.style={observed node,fill=white}
}
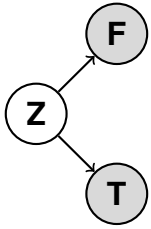
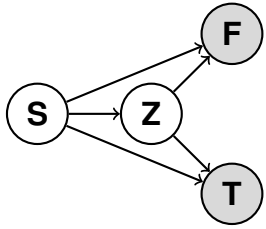
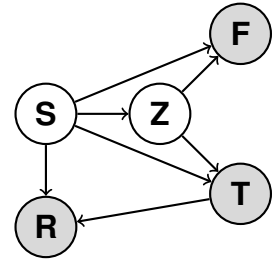
\begin{figure*}[!htb]
\centering
\begin{subfigure}[t]{0.25\textwidth}
\centering
\vskip 0pt 
\begin{tikzpicture}[->,node distance=1.5cm,thick]
  \node[unobserved node] (Z) {Z};
  \node[observed node] (F) [above right of=Z] {F};
  \node[observed node] (T) [below right of=Z] {T};

  \path
    (Z) edge (F)
    (Z) edge (T);
\end{tikzpicture}
\caption{Basic NMF (Section~\ref{sec:nmf})
\label{fig:nmf}}
\end{subfigure}
~
\begin{subfigure}[t]{0.35\textwidth}
\centering
\vskip 0pt 
\begin{tikzpicture}[->,node distance=1.5cm,thick]
  \node[unobserved node] (S) {S};
  \node[unobserved node] (Z) [right of=S] {Z};
  \node[observed node] (F) [above right of=Z] {F};
  \node[observed node] (T) [below right of=Z] {T};

  \path
    (S) edge (Z)
    (S) edge (F)
    (S) edge (T)
    (Z) edge (F)
    (Z) edge (T);
\end{tikzpicture}
\caption{NMF for source separation (Section~\ref{sec:nmf_ss})
\label{fig:nmf_ss}}
\end{subfigure}
~
\begin{subfigure}[t]{0.35\textwidth}
\centering
\vskip 0pt 
\begin{tikzpicture}[->,node distance=1.5cm,thick]
  \node[unobserved node] (S) {S};
  \node[observed node] (R) [below of=S] {R};
  \node[unobserved node] (Z) [right of=S] {Z};
  \node[observed node] (F) [above right of=Z] {F};
  \node[observed node] (T) [below right of=Z] {T};

  \path
    (T) edge (R)
    (S) edge (Z)
    (S) edge (R)
    (S) edge (F)
    (S) edge (T)
    (Z) edge (F)
    (Z) edge (T);
\end{tikzpicture}
\caption{Vibrato NTF (Section~\ref{sec:vib_ntf})
\label{fig:vibnmf}}
\end{subfigure}
\caption{Graphical models for the factorizations in this paper.
In each case the input data are a distribution over the observed (shaded) variables, while the model approximates the observation by a joint distribution over observed and latent (unshaded) variables that factorizes as specified.
$\mathsf{F}$, $\mathsf{T}$, $\mathsf{Z}$, $\mathsf{S}$, and $\mathsf{R}$ respectively represent the discrete frequencies, hops, components, sources, and frequency modulations over which the data is distributed.}
\end{figure*}

\subsection{Nonnegative matrix factorization}\label{sec:nmf}
We now summarize NMF within a probabilistic framework.
We consider the normalized Short-Time Fourier Transform (STFT) magnitudes (i.e., spectrogram) of the input signal as an observed discrete probability distribution of energy over the time-frequency plane, i.e.,
\begin{equation}\label{eq:spectrogram}
    p^{\text{obs}}(f, t) \triangleq \frac{|X(f, t)|}{\sum_{\nu, \tau}|X(\nu, \tau)|},
\end{equation}
$\forall f \in \{1, ..., F\}, t \in \{1, ..., T\} $, where $X$ is the input STFT and $(f, t)$ indexes the time-frequency plane.
NMF seeks an approximation $q$ to observed distribution $p^{\text{obs}}$ that is a valid distribution over the time-frequency plane and factorizes as
\begin{equation}\label{eq:nmf_factor}
    q(f, t) = \sum_{z}q(f|z)q(t|z)q(z) = \sum_{z}q(f|z)q(z, t).
\end{equation}
Figure \ref{fig:nmf} shows the graphical model for a joint distribution with this factorization.

We have introduced $z \in \{1, ..., Z\}$ as a latent variable that indexes components in the mixture, typically with $Z$ chosen to yield an overall data reduction, i.e., $FZ~+~ZT~\ll~FT$.
For a fixed $z_0$, $q(f|z_0)$ is a vector interpreted as the spectral template of the $z_0$-th component, i.e., the distribution over frequency bins of energy belonging to that component.
Likewise, $q(z_0, t)$ is interpreted as a vector of temporal activations of the $z_0$-th component, i.e., it specifies at what time indices the $z_0$-th component is prominent in the observed mixture.
Indeed, \eqref{eq:nmf_factor} can be implemented as a matrix multiplication, with the usual nonnegativity constraint on the factors satisfied implicitly, since $q$ is a valid probability distribution.

The optimization problem is typically formalized as minimizing the Kullback-Leibler (KL) divergence between the observation and approximation, or equivalently as maximizing the cross entropy between the two distributions:
\begin{equation}
\begin{aligned}
\underset{q}{\text{maximize}} & \quad \sum_{f, t} p^{\text{obs}}(f, t) \log q(f, t) \\
\text{subject to} & \quad q(f, t) = \sum_{z}q(f|z)q(z, t). \\
\end{aligned}
\end{equation}
While the non-convexity of this problem prohibits a globally optimal solution in reasonable time, a locally optimal solution can be found by multiplicative updates to the factors, which were first presented in \cite{lee01}.
We refer to this algorithm as KL-NMF, but note its equivalence to  Probabilistic Latent Component Analysis (PLCA) \cite{smaragdis06}, as well as a strong connection to topic modeling of counts data.

\subsection{NMF for source separation}\label{sec:nmf_ss}
NMF can be leveraged as a source model within a source separation task, such that the observed mixture is modeled as a sum of sources, each of which is modeled by NMF.
Whereas the latent variable $z$ in NMF indexes latent components belonging to a source, we now introduce an additional latent variable $s \in \{1, .., S\}$, which indexes latent sources within the mixture.
The resulting joint distribution over observed and latent variables is expressed as
\begin{equation}\label{eq:nmf_ss_joint}
q(f, t, s, z) = q(s)q(f|s, z)q(z, t|s).
\end{equation}
Thus the approximation to $p^{obs}(f, t)$ is the marginal distribution
\begin{equation}\label{eq:nmf_ss_approx}
\begin{aligned}
q(f, t) & = \sum_{s}q(s)q(f, t|s)\\
& = \sum_{s}q(s)\sum_{z}q(f|s, z)q(z, t|s),
\end{aligned}
\end{equation}
where $q(s_0)$ and $q(f, t|s_0)$ represent the mixing coefficient and NMF source model for the $s_0$-th source in the mixture, respectively.
Figure \ref{fig:nmf_ss} shows the graphical model.

Given a suitable approximation $q$, we estimate the latent sources in the mixture via Wiener filtering, i.e.,
\begin{equation}
X_s(f, t) = X(f, t) q(s|f, t),
\end{equation}
where the Wiener gains $q(s|f, t)$ are given by the conditional probabilities\footnote{A convenient result of the Wiener filter gains being conditional distributions over sources is that the mixture energy is conserved by the source estimates in the sense that $\sum_{s}X_s(f,t) = X(f,t) \medspace \forall \medspace f, t$.} of the latent sources given the approximating joint distribution
\begin{equation}
q(s|f, t) = \frac{q(f, t, s)}{q(f, t)} = \frac{\sum_{z}q(s)q(f|s, z)q(z, t|s)}{\sum_{z, s'}q(s')q(f|s', z)q(z, t|s')}.
\end{equation}
The estimated sources can then be reconstructed in the time-domain via the inverse STFT.

We seek a $q$ that both approximates $p^{\text{obs}}$ and yields source estimates $q(f, t|s)$ close to the true sources.
In a supervised setting, the spectral templates for each source model can be fixed by using basic NMF on some characteristic training examples in isolation.
When the appropriate training data is unavailable, the basic NMF can be extended by introducing priors on the factors or otherwise adding structure to the observation model to encourage, e.g., smoothness in the activations \cite{virtanen07} or harmonicity in the spectral templates \cite{bertin10}, which hopefully in turn improves the source estimates.
By contrast, our approach exploits local FM cues directly in the factorization, yielding an observation model for latent sources consistent with the sorts of pitch modulations expected in musical sounds.

\subsection{Coherent frequency modulation}
We now introduce frequency-slope-to-frequency ratios (FSFR) as local signal parameters under an additive sinusoidal model that are useful as grouping cues for the separation of sources with coherent FM, e.g. in the vibrato or glissando effects.
In continuous time, the additive sinusoidal model expresses the $s$-th source as a sum of component partials,\footnote{We do not assume any special structure in the partial frequencies, e.g., harmonicity.} each parameterized by an instantaneous frequency and amplitude, i.e.,
\begin{equation}
x_s(\tau) = \sum_{p=1}^{P}A_p(\tau) \cos \bigg( \theta_p(\tau_0) + \int_{\tau_0}^{\tau}\omega_p(u) du \bigg)
\end{equation}
where $p$ is the partial index, and $\theta_p(\tau_0)$, $A_p(\tau)$ and $\omega_p(\tau)$ specify the initial phase, instantaneous amplitude, and instantaneous frequency of the $p$-th partial.

We now consider a source under coherent FM, i.e.,
\begin{equation}\label{eq:cfm}
\omega_p(\tau) \triangleq (1+\kappa_s(\tau))\omega_p(\tau_0) \medspace \forall \medspace p
\end{equation}
for some modulation function $\kappa_s$ with $\kappa_s(\tau_0)=0$.
E.g., $\kappa_s$ resembles a slowly-varying sinusoid during frequency vibrato, or a gradual ramp function during glissando.
The FSFR are then expressed as
\begin{equation}\label{eq:fsfr}
\upsilon_p(\tau) \triangleq \frac{\omega_p'(\tau)}{\omega_p(\tau)} = \frac{\kappa_s'(\tau)}{1+\kappa_s(\tau)}.
\end{equation}
Note that $\{ \upsilon_p(\tau) \}$ are time-varying but independent of the partial index $p$ for a given source index $s$.
In other words, the instantaneous FSFR is common to all partials belonging to the same source and can be used as a grouping cue in unsupervised source separation \cite{creager15}.

\subsection{Distributed Derivative Method}
We now summarize the Distributed Derivative Method (DDM) \cite{betser09, hamilton12} for signal parameter estimation, which we use to estimate the FSFR at each time-frequency bin.
DDM estimates the parameters of a monochrome analytic signal under a $Q$-th order generalized sinusoid model,\footnote{
It is natural to specify the signal locally (near some time-frequency bin) as a generalized sinusoid even while the global model remains additive sinusoidal.
In particular, the notion of a time-frequency-localized signal follows from the filterbank summation interpretation of the STFT, and corresponds to the heterodyned and shifted input, prior to low-pass filtering by the window and downsampling in time \cite{allen77}.
In a slight abuse of notation, we later absorb the time-frequency indices as parameters in the analysis atom, i.e., we switch to the overlap-add interpretation of the STFT without warning.
} which is expressed as
\begin{equation}\label{eq:gen_sine}
x(\tau) = \exp \bigg( \sum_{q=0}^{Q}\eta_{q}\tau^{q} \bigg),
\end{equation}
where $\boldsymbol{\eta} \in \mathbb{C}^{Q+1}$ is the vector of signal parameters, whose real and imaginary parts specify the log amplitude law and phase law,\footnote{The frequency law is trivially computed from the phase law.} respectively.
In this work, we specify \eqref{eq:gen_sine} as a constant amplitude signal with linear frequency modulation, i.e., $\boldsymbol{\eta} \in \mathbb{C}^{3}$ with $\Re (\eta_i )= 0~\,~\forall~\,~i$.
The signal parameters $\Im (\eta_1 )$ and $\Im (\eta_2 )$ then specify (within multiplicative constants) the instantaneous frequency and frequency slope, respectively.

The parameters of interest can be estimated by considering the inner product of the signal with a family of differentiable analysis atoms of finite time-frequency support.
In particular, the continuous-time STFT can be expressed by inner product as
\begin{equation}\label{eq:cstft}
\X(f, t) \triangleq \langle x(\tau), \phi(\tau; f, t) \rangle = \int_{\tau=-\infty}^{+\infty} x(\tau)\phi(\tau; f, t)^{*} d\tau,
\end{equation}
where $\X(f, t)$ is the STFT, $x(\tau)$ is the input signal, and $\phi(\tau; f, t)$ is a heterodyned window function from some differentiable family (e.g. Hann), parameterized by its localization $(f, t)$ in the time-frequency plane.
The signal parameters are solutions to equations of the form
\begin{equation}\label{eq:ddm}
\langle x(\tau), \phi'(\tau; f, t) \rangle = - \sum_{q=1}^{Q} \eta_q \langle q \tau^{q-1} x(\tau), \phi(\tau; f, t) \rangle,
\end{equation}
which is linear in $\lbrace \eta_q \rbrace$ for $q > 0$, and permits an STFT-like computation of both inner products.
The right-hand side of \eqref{eq:ddm} is derived from the left-hand side using integration by parts, exploiting the finite support of $\phi(\tau; f, t)$, and substituting in the signal derivative $x'(\tau)$ from \eqref{eq:gen_sine}.
To estimate the signal parameters at a particular $(f_0, t_0)$, we construct a system of linear equations by evaluating \eqref{eq:ddm} for each $\phi(\tau; f, t)$ in a set of nearby atoms $\Phi$, then solve for $\boldsymbol{\eta}$ in a least-squares sense.
We typically use atoms in neighboring frequency bins at the same time step, i.e., $\Phi~=~\lbrace \phi(\tau; t_0, f_0-\frac{L-1}{2}),~...,~\phi(\tau; t_0, f_0+\frac{L-1}{2})\rbrace$ for some odd $L$.

While DDM is an unbiased estimator of the signal parameters in continuous time, we must implement a discrete-time approximation on a computer.
This introduces a small bias that can be ignored in practice since the STFT window is typically longer than a few samples \cite{betser09}.
\section{Proposed Method}\label{sec:vib_nmf}
\begin{figure*}[!htb]
\centering
\begin{subfigure}[t]{0.4\textwidth}
\centering
\vskip 0pt 
\includegraphics[width=1.0\textwidth,natwidth=610,natheight=642]{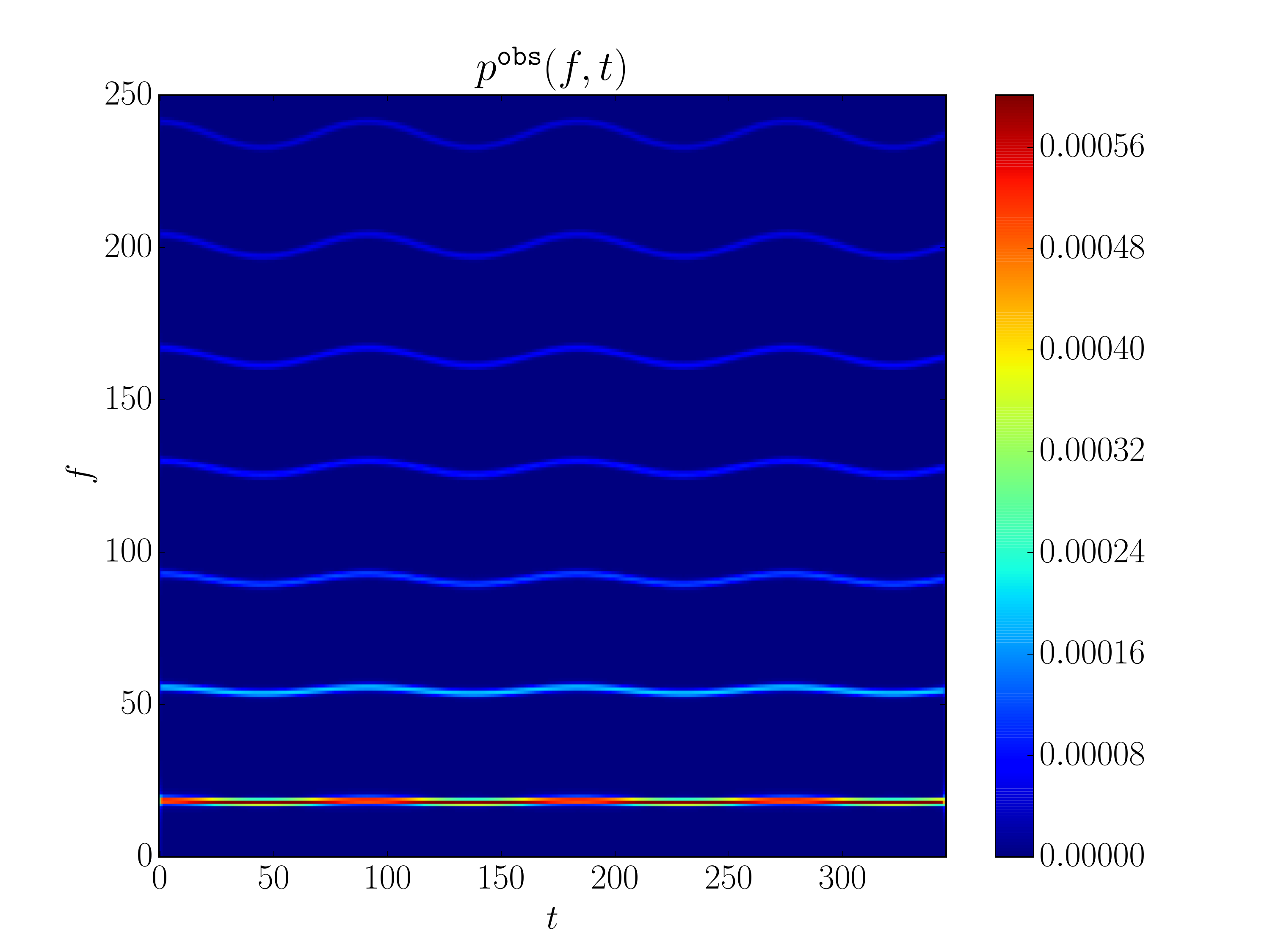}
\caption{Spectrogram}
\label{fig:mags}
\end{subfigure}
\qquad
~
\begin{subfigure}[t]{0.4\textwidth}
\centering
\vskip 0pt 
\includegraphics[width=1.0\textwidth,natwidth=610,natheight=642]{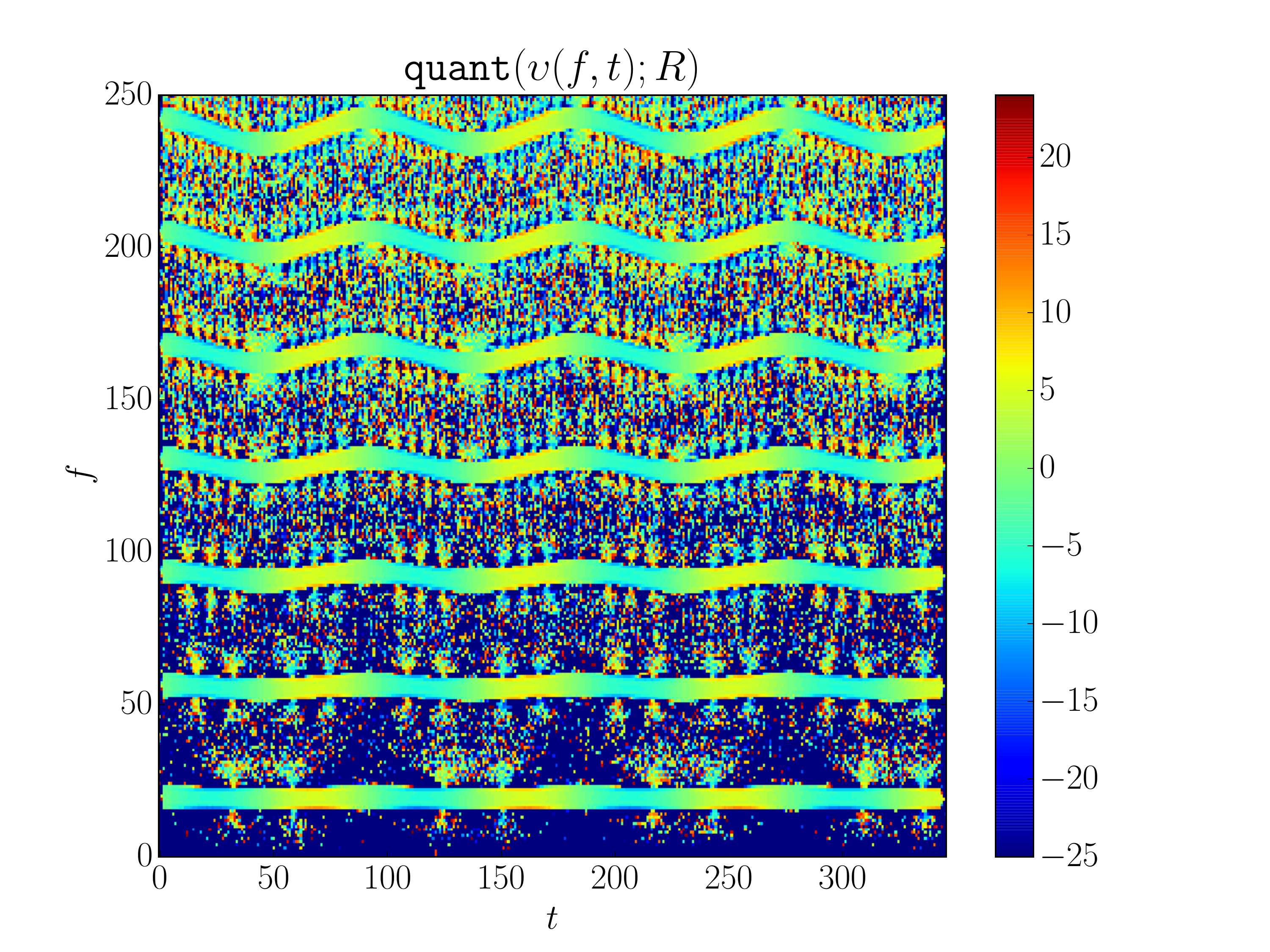}
\caption{Discretized FSFR}
\label{fig:fsfr}
\end{subfigure}
\caption{Unfolding the nonzero elements in the observation tensor for a synthetic vibrato square wave note (G5).
The hop index $t$ spans 2 seconds of the input audio, while the bin index $f$ spans half the sampling rate, 0--22.05 kHz.}
\label{fig:tensor}
\end{figure*}

\subsection{Motivation}
The NMF signal model is not sufficiently expressive to compactly represent a large class of musical sounds, namely those characterized by slow frequency modulations, e.g., in the vibrato effect.
In particular, it specifies a single fixed spectral template per latent component and thus requires a large number of components to model sounds with nonstationary pitch.
From a separation perspective, as the number of latent components grows, so grows the need for a comprehensive model that can correctly group components belonging to the same source.
To this end, we appeal to the perceptual theory of Auditory Scene Analysis\cite{bregman90}, which postulates the importance of shared frequency or amplitude modulations among partials as a perceptual cue in their grouping \cite{chowning80, mcadams89}.
In this work we focus on FM, although in principle our approach could be extended to include amplitude modulations.\footnote{In turn, this would increase the dimensionality of the data.}
We now propose an extension to KL-NMF that leverages this so-called common fate principle and is suitable for the analysis of vibrato signals.

\subsection{Compiling the observations as a tensor}
DDM yields the local estimates of frequency and frequency slope for each time-frequency bin, from which the FSFR are trivially computed.
We define the (sparse) observation tensor
$\p \in \mathbb{R}^{F \times T \times R}_{\geq 0}$
as an assignment of the normalized spectrogram into one of $R$ discrete bins for each $(f, t)$ according the local FSFR estimate, i.e.,
\begin{equation}
\p \triangleq \begin{cases}
    p^{\text{obs}}(f, t) & \text{if} \quad \texttt{quant}(\upsilon(f, t); R) = r \\
    0 & \text{else},
    \end{cases}
\end{equation}
where $p^{\text{obs}}(f, t)$ is the normalized spectrogram as in \eqref{eq:spectrogram} and $\upsilon$ are the FSFR as in \eqref{eq:fsfr}, which are quantized by $\texttt{quant}(\cdot; R)$, possibly after clipping to some reasonable range of values.
Figure \ref{fig:tensor} shows the spectrogram and FSFR for a synthetic vibrato square wave.
\begin{figure*}[!htb]
\centering
\begin{subfigure}[t]{0.4\textwidth}
\centering
\vskip 0pt 
\includegraphics[width=1.0\textwidth,natwidth=610,natheight=642]{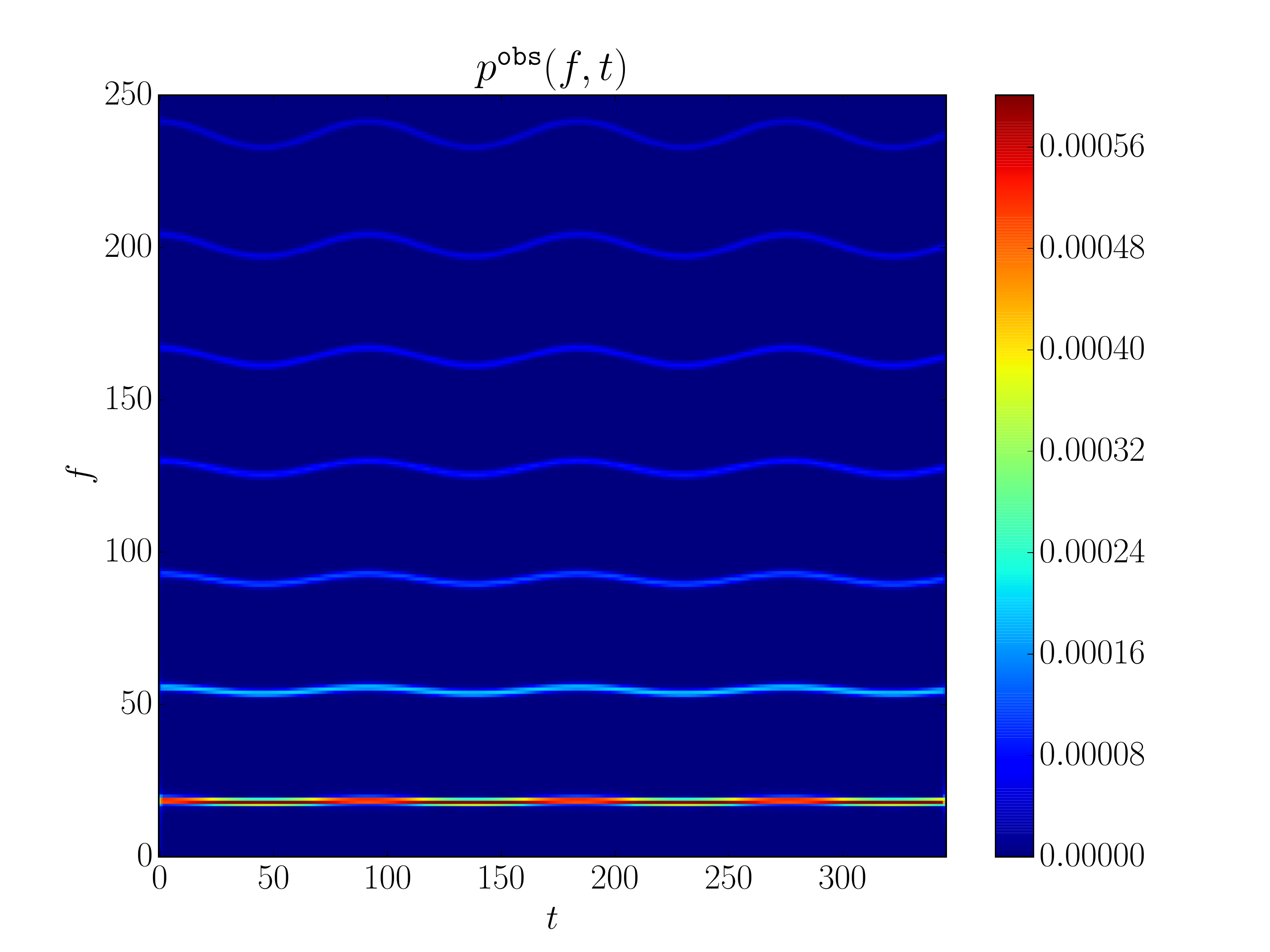}
\caption{Spectrogram}
\label{fig:mags_synth}
\end{subfigure}
~
\begin{subfigure}[t]{0.4\textwidth}
\centering
\vskip 0pt 
\includegraphics[width=1.0\textwidth,natwidth=610,natheight=642]{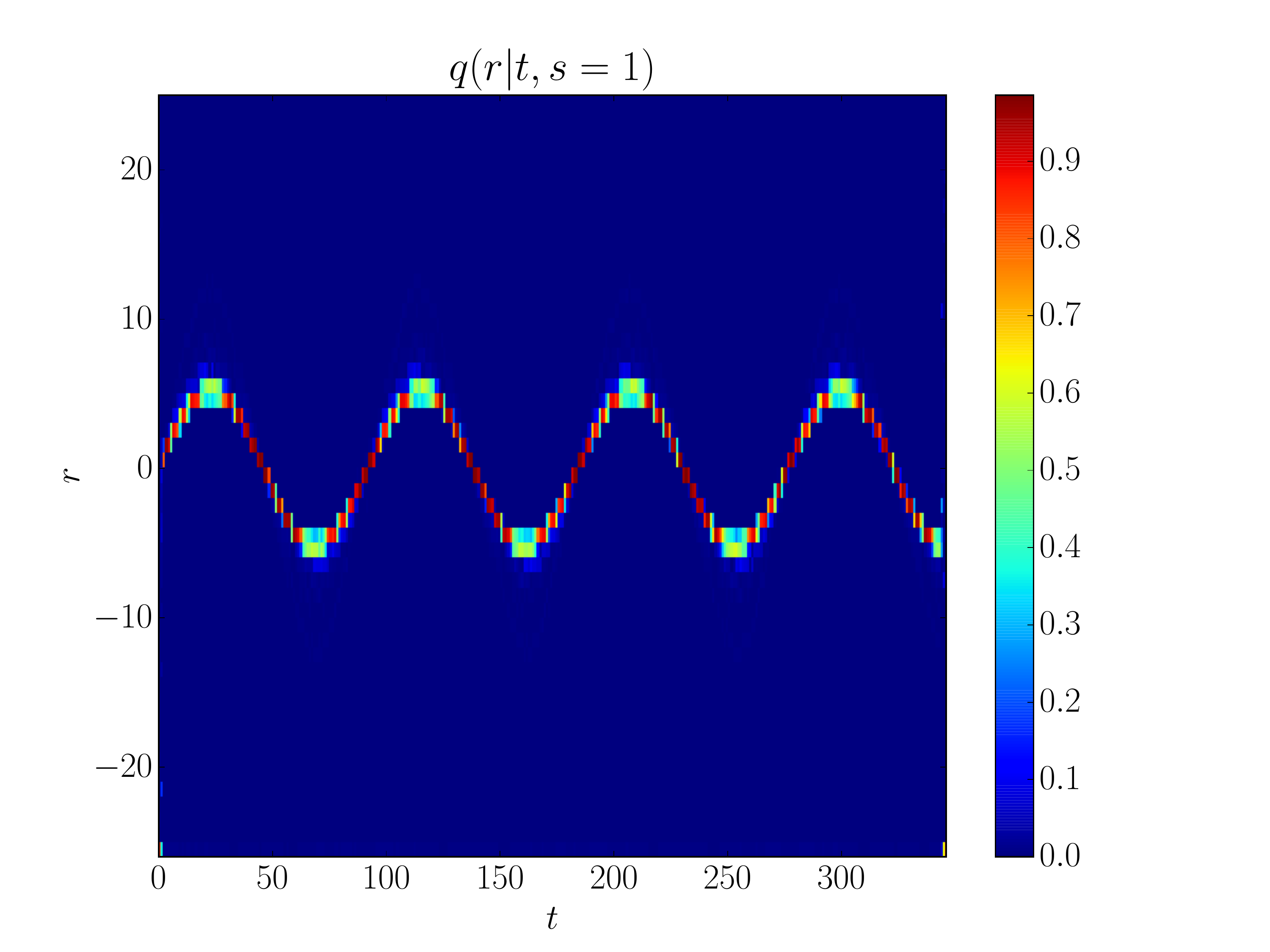}
\caption{VibNTF FM model}
\label{fig:fsfr_synth}
\end{subfigure}
\\
\begin{subfigure}[t]{0.4\textwidth}
\centering
\vskip 0pt 
\includegraphics[width=1.0\textwidth,natwidth=610,natheight=642]{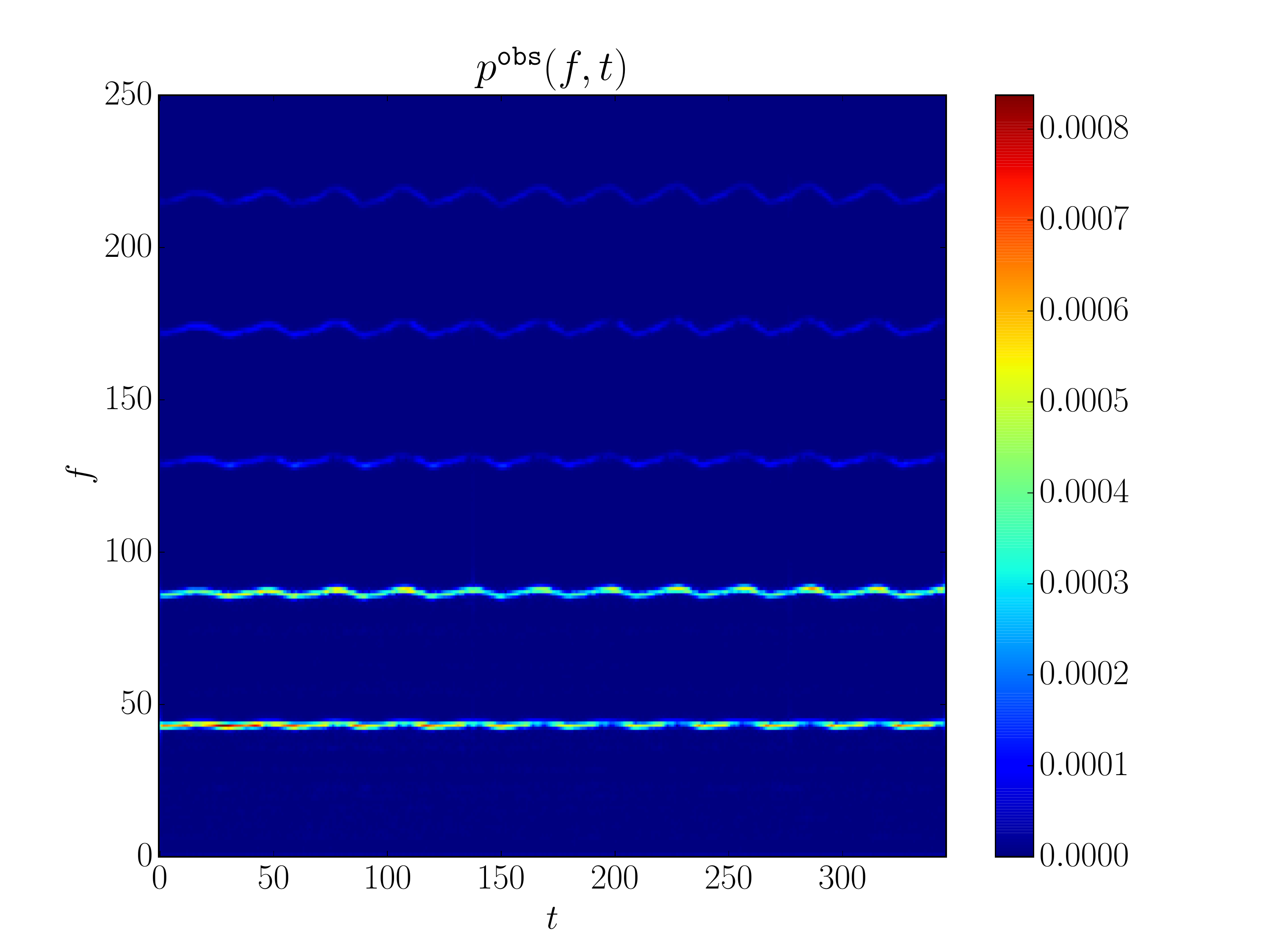}
\caption{Spectrogram}
\label{fig:mags_real}
\end{subfigure}
~
\begin{subfigure}[t]{0.4\textwidth}
\centering
\vskip 0pt 
\includegraphics[width=1.0\textwidth,natwidth=610,natheight=642]{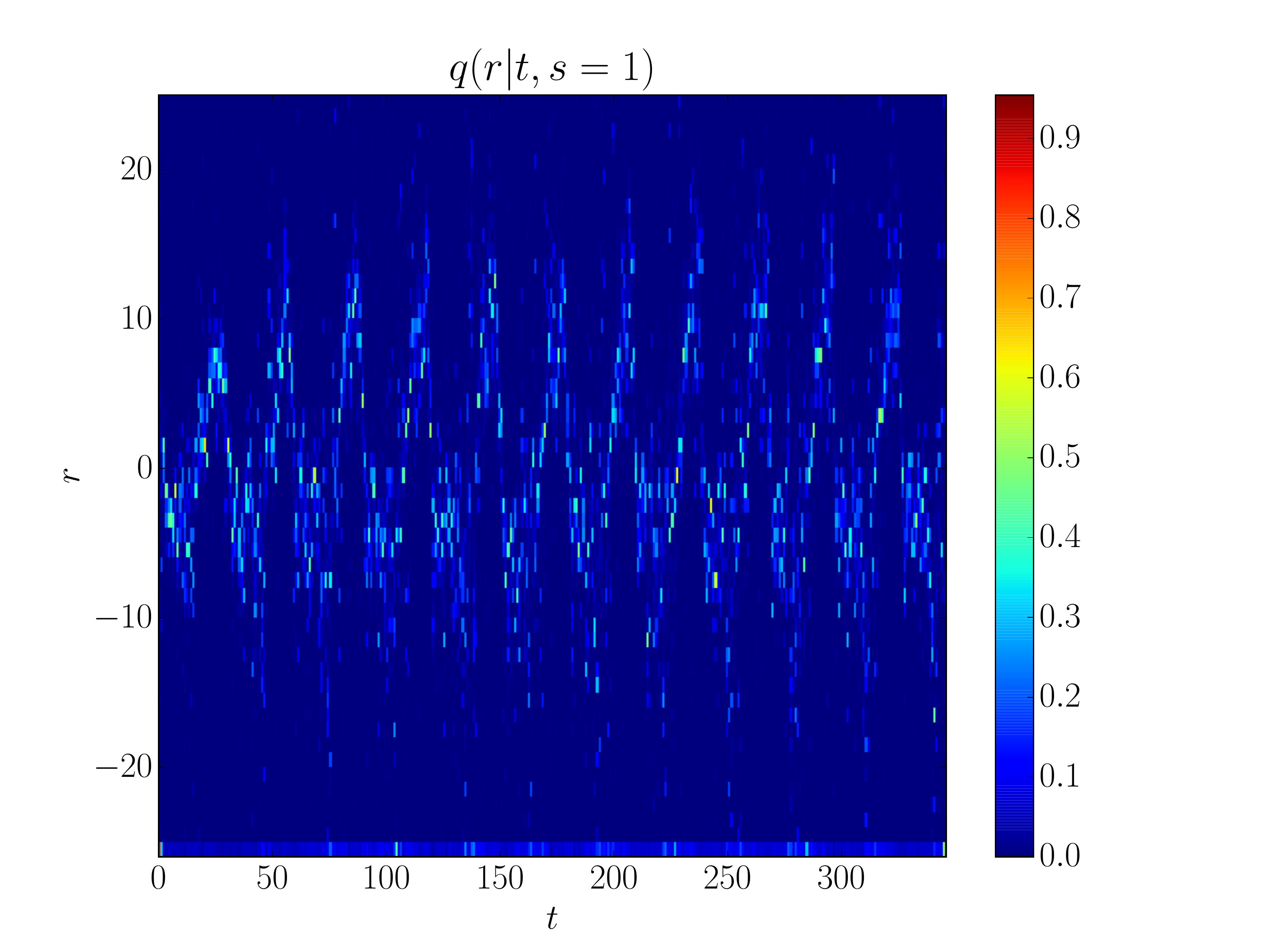}
\caption{VibNTF FM model}
\label{fig:fsfr_real}
\end{subfigure}

\caption{For single-note analyses, VibNTF encodes the time-varying pitch modulation.
The top row shows a synthetic vibrato square wave note (G5), while the bottom row shows a real recording of a violin vibrato note ($\text{B}^\flat$6).
We plot $r$ in the range $[-\frac{R}{2}, \frac{R}{2}]$ in figures \ref{fig:fsfr_synth} and \ref{fig:fsfr_real} to clarify that the index $r$ represents a zero-mean quantity (the FSFR).}
\label{fig:singlenote}
\end{figure*}

\subsection{Vibrato NTF}\label{sec:vib_ntf}
As with NMF, we seek a joint distribution $q$ with a particular factorized form, whose marginal maximizes cross entropy against the observed data.
We propose an observation model of the form
\begin{equation}\label{eq:approx}
q(f, t, \fsfr) = \sum_{s}q(s)q(\fsfr | t, s)\sum_{z}q(f|s, z)q(z, t|s)
\end{equation}
where $q(s)$ represents the mixing, $q(\fsfr | t, s)$ represents the common time-varying FSFR per source, and $\sum_{z}q(f|s, z)q(z, t|s)$ represents the NMF source model.
Figure \ref{fig:vibnmf} shows the graphical model of the joint distribution.
Thus, given $p^{\text{obs}}$ we seek an approximation $q$ that factorizes as in \eqref{eq:approx} and maximizes
\begin{equation}\label{eq:ce}
\begin{aligned}
    \alpha(q) & \triangleq \sum_{f, t, \fsfr} \p \log q(f, t, \fsfr) \\
    & = \sum_{f, t, \fsfr} \p \log \sum_{z, s} q(f, t, \fsfr, z, s).
\end{aligned}
\end{equation}
The sum in the argument to the log makes this difficult to solve outright, so we find a local optimum by iterative Minorization-Maximization (MM) \cite{hunter04} instead.
That is, given $q^{(i)}$, our model at the current ($i$-th) iteration, we pick a better $q^{(i+1)}$ by (a) finding a concave minorizing function $\beta(q; q^{(i)})$ such that $\beta(q; q^{(i)}) \leq \alpha(q) \medspace \forall \medspace q$
and $\beta(q^{(i)}; q^{(i)}) =\alpha(q^{(i)})$,
 and (b) maximizing $\beta(q; q^{(i)})$ with respect to $q$.

In particular, $\beta(q; q^{(i)})$ is derived\footnote{Cf. \cite{stein15} for a more thorough treatment.} by applying Jensen's inequality to \eqref{eq:ce}, and is expressed as
\footnotesize
\begin{equation}\label{eq:beta}
\beta(q; q^{(i)}) \triangleq \sum_{f, t, \fsfr, z, s} \p q^{(i)}(z, s | f, t, \fsfr) \log \frac{q(f, t, \fsfr, z, s) }{q^{(i)}(z, s | f, t, \fsfr)},
\end{equation}
\normalsize
where $q^{(i)}(z, s | f, t, \fsfr)$ is the approximate posterior over latent variables given the model at the $i$-th iteration\footnote{Note that the MM iteration specifies an expectation-maximization.}, computed as
\begin{equation}
q^{(i)}(z, s | f, t, \fsfr) = \frac{q^{(i)}(z, s, f, t, \fsfr)}{\sum_{z',s'}q^{(i)}(z', s', f, t, \fsfr)}.
\end{equation}

For notational convenience we define $\rho(f, t, \fsfr, z, s)~\triangleq~\p q^{(i)}(z, s | f, t, \fsfr)$ and discarding the denominator in the log of \eqref{eq:beta} (constant w.r.t.\ $q$), equivalently write the optimization over the minorizing function as
\small
\begin{equation}\label{eq:opt}
\underset{q}{\operatorname{max}} \sum_{f, t, \fsfr, z, s} \rho(f, t, \fsfr, z, s) \log q(s) q(f|z, s) q(z, t|s)q(\fsfr| t, s).
\end{equation}
\normalsize
We now alternatively update each factor by separating the argument in the log in \eqref{eq:opt} as a sum of logs, each term of which can be optimized by applying Gibb's inequality\cite{mackay03}.
That is, given the current model, the optimal choice for some factor of $q^{(i+1)}$ is the marginal of $\rho$ over the corresponding variables.
E.g.,
\begin{subequations}
\begin{equation}
\begin{aligned}
q^{(i+1)}(s) & \leftarrow \frac{\sum_{f, t, \fsfr, z} \rho(f, t, \fsfr, z, s)}{\sum_{f, t, \fsfr, z,s'} \rho(f, t, \fsfr, z, s')}.
\end{aligned}
\end{equation}
Likewise, the remaining factor updates are expressed as
\begin{equation}
q^{(i+1)}(f|z, s) \leftarrow \frac{\sum_{t,r} \rho(f,t,r,z,s)}{\sum_{f',t,r} \rho(f',t,r,z,s)};
\end{equation}
\begin{equation}
q^{(i+1)}(z, t| s) \leftarrow \frac{\sum_{f, \fsfr} \rho(f,t,\fsfr,z,s)}{\sum_{f, t',\fsfr,z'} \rho(f,t',\fsfr,z',s)};
\end{equation}
\begin{equation}
q^{(i+1)}(\fsfr| t, s) \leftarrow \frac{\sum_{f, z} \rho(f,t,\fsfr,z,s)}{\sum_{f,\fsfr',z} \rho(f,t,\fsfr',z,s)}.
\end{equation}
\end{subequations}
Since $\rho$ is expressed as a product of the current factors and observed data, the factor updates can be implemented efficiently by using matrix multiplications to sum across inner dimensions as necessary.
The theory guarantees convergence\footnote{For guaranteed convergence, $\rho$ must be recomputed after each factor update, rather than once per iteration as the notation suggests.
However, in practice we observe convergence without the recomputation.} to a local minimum \cite{hunter04}, although in practice we stop the algorithm after some fixed number of iterations.
The algorithm is initialized by choosing factors of $q^{(0)}$ as random valid conditional probabilities.

Figure \ref{fig:singlenote} visualizes the FM factor $q(\fsfr|t, s)$ estimated by the proposed algorithm for single note analyses ($S=1$) of both synthetic and real data.
\section{Evaluation}\label{sec:results}
We present a comparison of our proposed method with the baseline KL-NMF (which our method extends) in a blind source separation task examining mixtures of two single-note recordings.
We use the $\texttt{BSS\_EVAL}$ criteria~\cite{vincent06} to evaluate separation performance, which necessitates the use of artificial mixtures.
We report the source-to-distortion ratio (SDR), source-to-interference ratio (SIR), and source-to-artifact ratio (SAR), each in dB.
Each experiment comprises 500 separations, with the sources in each trial chosen as specified below and mixed at $0$ dB with a total mixture duration of 2 seconds at 44.1 kHz sampling rate.
We report the average metrics across all sources and trials.

To use KL-NMF for blind source separation, we must specify $Z=2$, i.e., each mixture component considered as a source.
This baseline should be relatively easy to beat, since empirically KL-NMF does a poor job of modeling vibrato signals when $Z$ is small.

For Vibrato NTF, we specify $S=2$ and $Z=3$, i.e., for each of the two sources we learn spectral templates and temporal activations for three components.
E.g., considering a sinusoidal vibrato, the components could model the source during the crest, midpoint, and trough of the pitch modulation.
We estimate the signal parameters at a particular $(f_0, t_0)$ using DDM with a family of $L=5$ analysis atoms (heterodyned Hann functions) in the same hop index and nearby frequency bins.
In order to avoid the influence of noisy FSFR estimates in the factorization, we apply some mild post-processing prior to quantization.
Specifically, we implicitly discard FSFR at $(f, t)$ with $p^{\text{obs}}(f,t)$ below the 10$^{\text{th}}$ percentile, or outside a reasonable range of $\pm$4 times the sampling rate by setting them to the data median.
The FSFR are then quantized evenly across their range into $R=50$ discrete values.
\begin{table}[tb]
\centering
\begin{tabular}{m{2.5cm} m{1.5cm} m{1.5cm} m{1.5cm}}
\toprule
&\multicolumn{3}{c}{\texttt{BSS\_EVAL} in dB} \\
Algorithm & SDR & SIR & SAR\\
\midrule
\emph{(A)~Synthetic~data} & & & \\
2-part KL-NMF&-1.5~$\pm$~0.1&0.1~$\pm$~0.2&6.9~$\pm$~0.2\\
Vibrato NTF&14.6~$\pm$~1.0&17.0~$\pm$~1.2&23.6~$\pm$~0.7\\
\emph{(B)~Real~data} & & & \\
2-part KL-NMF&2.8~$\pm$~0.4&8.0~$\pm$~2.1&9.2~$\pm$~0.2\\
Vibrato NTF&5.8~$\pm$~0.5&9.7~$\pm$~2.2&17.7~$\pm$~0.5\\
\bottomrule
\end{tabular}
\caption{Mean and 95\% confidence intervals of the $\texttt{BSS\_EVAL}$ metrics for 500 unsupervised separations of two-source mixtures.
Experiment A considers synthetic vibrato square waves, while experiment B considers single-note vibrato string instrument recordings.}
\label{tab:results}
\end{table}

For both algorithms, the STFT in \eqref{eq:spectrogram} is specified by a 1024-length (23 msec) Discrete Fourier Transform using a Hann window with 75\% overlap between successive frames.
Thus, $F=513$, corresponding to the non-redundant frequency bins, and $T=346$, the number of hops required to cover the mixture duration.
Both algorithms are initialized randomly and run for $100$ iterations.

Experiment A examines synthetic data, where the sources are square waves with frequency vibrato, whose signal parameters are generated at random.
The fundamental frequency corresponds to a note value selected uniformly at random from the three-octave range [A3, $\text{G}^\sharp$5].
The number of partials is chosen uniformly at random from the range [10, 30], and subsequently reduced as necessary to avoid aliasing.
The vibrato modulation function, i.e., $\kappa_s$ in \eqref{eq:cfm}, is a sinusoid with depth chosen uniformly at random in the range of [5\%, 20\%] of the fundamental and rate chosen log-uniformly at random from the range [0.5, 10] Hz.

Experiment B examines real data, where the sources are single-note recordings from the McGill University Master Samples (MUMS)~\cite{opolko87}, which contains over 6000 single-note and single-phrase recordings of classical and popular instruments.
We focus our evaluation on string instruments, which exhibit strong frequency modulation in their vibrato effect~\cite{verfaille05}.
The MUMS subset of string instrument notes with vibrato comprises a total of 250 unique recordings of violin, viola, cello, and double bass.
The sources are chosen randomly from this subset and trimmed or padded to 2 seconds as necessary.

Results for both experiments are provided in table \ref{tab:results}.
Experiment A shows a dramatic win for Vibrato NTF over the baseline.
We see some variability in the results, which reflects an optimization over a cost surface with many local optima.
With random initialization, Vibrato NTF works either very well or very poorly, so robustness could be improved by a more careful initialization, or alternatively by regularizing the factorization in such a way as to avoid suboptimal solutions.

In experiment B, we see that moving from synthetic to real data degrades the performance of our proposed method, although we still beat the baseline by a modest margin.
Interestingly, the baseline performs better on real data than synthetic, likely because the pitch variations are less pronounced so KL-NMF fails less frequently.
Moreover, the pitch modulations in real data are more complex than in the synthetic case (compare figures \ref{fig:fsfr_synth} and \ref{fig:fsfr_real}), and may require more components (larger $Z$) to be properly modeled.
Vibrato NTF as proposed tends to decrease in performance as $Z$ increases, so additional work is required to improve robustness for the analysis of real data.
We hypothesize that an extension enforcing temporal continuity in the FM factor, which should be smooth and monotonic per-source, would enhance the grouping of components, permitting a larger $Z$ in practice.
\section{Conclusion}\label{sec:conclusion}
We proposed Vibrato NTF, a novel blind source separation algorithm that extends NMF by leveraging local estimates of frequency modulation as grouping cues directly in the factorization.
Experimental results using synthetic data showed a substantial improvement over the baseline, and validated the FSFR as useful grouping cues in a source separation task.
In the experiment with real recordings, our method provided a more modest improvement.
With regards to the analysis of real data, we believe the incorporation of sensible priors on the factors would improve the separation performance, while careful initalization would improve the robustness.
Further work could include tailoring the proposed method to the analysis of polyphonic sounds, or sounds with mild or no frequency modulation.
Additionally, an extension including coherent amplitude modulations as a grouping cue is possible within the proposed tensor factorization framework.
\section{Acknowledgements}\label{sec:acknowledgements}
The research leading to this paper was partially supported by the French National Research Agency (ANR) as a part of the EDISON 3D project (ANR- 13-CORD-0008-02), and by the Canadian National Science and Engineering Research Council (NSERC).
Additional support was provided by the Analog Garage, the emerging business accelerator at Analog Devices, Inc.
\newpage
\clubpenalty10000
\widowpenalty10000
\bibliography{./vib}
\bibliographystyle{plainurl}
\end{document}